\documentclass{article}
\usepackage{amsmath, amssymb, latexsym}
\usepackage{braket}
\usepackage[dvips]{graphicx}
\usepackage{url}
\usepackage{cite}
\usepackage{amscd}
\usepackage[normalem]{ulem}

\newcommand{\Slash}[1]{{\ooalign{\hfil/\hfil\crcr$#1$}}}

\begin{document}

\renewcommand{\thefootnote}{\fnsymbol{footnote}}

\begin{flushright}
\hfill UT-KOMABA/14-1
\newline
\hfill KEK-TH-1737
\end{flushright}

\vspace*{2mm}

\begin{center}
{\bf{\large Finite pulse effects on $e^+e^-$ pair creation 
from strong electric fields}}

\vspace*{10mm}

H.~Taya$^{(a,b)}$\footnote[1]{\url{h_taya@hep1.c.u-tokyo.ac.jp}}, 
H.~Fujii$^{(a)}$\footnote[2]{\url{hfujii@phys.c.u-tokyo.ac.jp}}, 
and 
K.~Itakura$^{(c,d)}$\footnote[3]{\url{kazunori.itakura@kek.jp}}\\

\vspace*{5mm}

$^{(a)}$ {\it Institute of Physics, University of Tokyo, Komaba 3-8-1, 
Tokyo 153-8902, Japan}\\
$^{(b)}$ {\it Department of Physics, The University of Tokyo, 
Hongo 7-3-1, Bunkyo-ku, Tokyo 113-0033, Japan}\\
$^{(c)}$ {\it Theory Center, IPNS, 
High Energy Accelerator Research Organization 
(KEK), Tsukuba, Oho 1-1, Ibaraki 305-0801, Japan}\\
$^{(d)}$ {\it Department of Particle and Nuclear Studies, Graduate University for Advanced Studies (SOKENDAI), Oho 1-1, Tsukuba, Ibaraki 305-0801, Japan}

\vspace*{3mm}
\end{center}

\begin{abstract}
We investigate electron-positron pair creation 
from the vacuum in a pulsed electric background field.
Employing the Sauter-type pulsed field
$E(t)=E_0 {\rm sech}^2 (t/\tau)$ with height $E_0$ and
width $\tau$, we demonstrate 
explicitly the interplay between the nonperturbative 
and perturbative aspects of pair creation
in the background field.  
We analytically compute the number of produced pairs 
from the vacuum in the Sauter-type
field, and the result reproduces Schwinger's nonperturbative formula
in the long pulse limit (the constant field limit),
while in the short pulse limit it coincides with the leading-order 
perturbative result.
We show that two dimensionless parameters 
$\nu = |eE_0| \tau^2$ and $\gamma = |eE_0| \tau /m_e$
characterize the importance of multiple interactions 
with the fields and the transition from the perturbative to the 
nonperturbative regime.  We also find 
that pair creation is enhanced compared to Schwinger's formula 
when the field strength is relativity weak $|eE_0|/m_e^2 \lesssim 1$ and 
the pulse duration is relatively short $m_e \tau \lesssim 1$,
 and reveal that the enhancement is predominantly described 
by the lowest order perturbation with a single photon. 
\end{abstract}

\vspace*{5mm}

\section{INTRODUCTION}


In the presence of extraordinarily strong gauge fields, we encounter
essentially new phenomena that are not observed in the vacuum. Such
phenomena are collectively called ``strong-field physics," 
which has been attracting the attention of many researchers 
in various fields in physics \cite{PIF}. 
There are intense laser facilities planned around the world; in addition, 
compact stars and relativistic nucleus-nucleus collisions
offer unique opportunities to study the strong-field phenomena.
Even for a system with a small coupling constant,
the physics becomes nonperturbative because
the smallness of the coupling constant is compensated 
by the strong background field strength, 
which requires some sort of resummations of higher-order contributions.
Such nonperturbative nature of interactions 
between particles and fields is one of the
outstanding properties of strong-field physics. 
At the same time,
it is an important issue to understand the transition 
from the perturbative to the nonperturbative regime 
with increasing the strength of the fields
and/or with changing other parameters.
This is relevant also to experiments for realizing the setups
for strong-field physics.
The present paper is devoted to clarifying the
interplay between perturbative and nonperturbative aspects of the
phenomena under strong fields.

To be more specific, let us consider a system described by QED in a
very strong electric field $|eE| \gtrsim m^2_e$ where $m_e$ is the
electron mass \cite{StrongQED}. In such a system, propagation of an electron
significantly differs from that of the bare one because it receives 
large corrections from the strong field.  
From a naive dimensional argument, 
the propagation of an electron 
acquires corrections of the order of
${\mathcal O}([eE/m_e^2]^n)$ 
if it receives $n$ kicks from the strong field. 
Therefore, for $|eE| \gtrsim m^2_e$,
the higher-order interactions are not suppressed and 
the propagation of an electron bears nonperturbative nature.

Typical examples of the nonperturbative phenomena include the
Schwinger mechanism \cite{sau31,EH36,sch51}, i.e., spontaneous
production of $e^+ e^-$ pairs from the vacuum (see Ref.~\cite{Dunne}
for a review). Intuitively it is understood as the process where 
a virtual $e^+ e^-$ pair produced 
as a vacuum fluctuation is kicked many times by the strong
field so that they obtain enough energy to become a real pair.
Formally, 
an imaginary part appears in the effective action of the background
fields (the Euler-Heisenberg action \cite{EH36}) only after summing up
the electron's one-loop diagrams with 
infinitely many insertions of the external fields. 
The total number of created electrons in
case of a constant electric background field with infinite duration is
given by $N/(TV) \propto |eE|^2 \exp\{ -\pi m_e^2/|eE| \}$ whose 
dependence on $eE$ 
clearly indicates nonperturbative nature 
of the Schwinger mechanism \cite{sch51}.

Another aspect of strong fields which becomes important 
in actual physical situations
is the temporal dependence of the fields. 
As far as a strong electric field is concerned, it is not realistic
to keep it for a long time compared to the typical time scale of the
system. 
For example, electric fields produced in high-energy heavy-ion
collisions decay quite fast.
Those in intense lasers are also time dependent. 
Based on the intuitive picture that charged particles
receive quantum corrections by kicks from the strong fields, one may
expect that the number of kicks from the fields depends on 
the field lifetime.
Indeed, perturbative calculation may be justified 
for a short-lived pulse field like a shock wave or a spike.
Thus, we need to study carefully 
whether the process should be described in a perturbative or nonperturbative way for the pair creation
in the presence of time-dependent strong fields.
This is the problem we are going
to address in the present paper.

Before going into the details, let us briefly explain here 
how the finite duration of the fields modifies
our intuitive picture for the pair creation in the background
field.
In a static and homogeneous electric field, 
the system has only
 two dimensionful parameters,
$eE$ and $m_e$. As mentioned above, a criterion for pair creation from
the vacuum is then 
given by $|eE|/m_e^2\gtrsim 1$. This is also understood in
the following way.  Pair creation will become possible if the work,
$|eE|d$, done by the field on a virtual electron/positron
for a distance $d$ is at least
of the same order as the electron mass. Since this must happen within
the lifetime of pair fluctuation, the distance $d$ 
is identified with the Compton length $d\sim 1/m_e$. 
However, 
when the external electric field is time dependent, 
 we have another dimensionful parameter $\tau$, namely, 
a typical duration of the field. In the static field case, we care 
only the lifetime of the fluctuation, but now we need to 
deal with the two time scales: $\tau$ and $1/m_e$. 
We will find that two dimensionless parameters 
$\gamma=|eE_0| \tau/m_e$ and $\nu=|eE_0|\tau^2$ are relevant for this
time dependent case, and can discuss the interplay between perturbative
and nonperturbative physics with the values of $\gamma$ and $\nu$.
In particular, we can study the interplay in detail by using the 
Sauter-type electric
field $E(t)=E_0{\rm sech}^2(t/\tau)$ which allows analytic calculation
for the Schwinger mechanism\footnote{The parameter dependence of the pair creation in the Sauter-type electric field was studied previously by solving quantum kinetic equations numerically\cite{heb08, koh13, SkokovLevai09, sko07}.  }.

The present paper is organized as follows: In the next section, we
provide perturbative and nonperturbative formulations for the
$e^+e^-$ pair creation in a time-dependent electric field without
specifying any temporal profile. In section 3, we take the Sauter-type
field as an example of pulsed electric fields and compute the number
densities of produced electrons in both formulations. Then, we compare
the two results. A summary is given in the last section.

\section{PAIR CREATION IN TIME-DEPENDENT ELECTRIC FIELDS}
\label{sec2}

The purpose of this section is to present a general 
expression for the number of electrons created from the 
vacuum in the presence of a time-dependent electric field. 
To this aim, we consider the following QED Lagrangian: 
$
{\mathcal L} 
       = \bar{\psi} \left[  i\Slash{\partial} 
    -m   \right]\psi   - e\bar{\psi}\Slash{\bar{A}} \psi 
	\equiv {\mathcal L}_0 + {\mathcal L}_{\rm BG} \, .
$
Here $e>0$ is the coupling constant, $\psi$ is 
the electron field, 
and $\bar{A}^{\mu}$ is the background gauge field.  We assume a background electric field \mbox{\boldmath $E$} directed to 
the $z$ axis, which is homogeneous in space but depends on time,
and we set the background gauge field four-potential 
$\bar{A}^{\mu}$ to
\begin{align}
	\bar{A}^{\mu}(x) = (0, 0, 0, -\int^t_{-\infty} E(t') dt')\, .  
\end{align}
Below we first derive a formula for the number of produced 
electrons in the lowest-order perturbation theory, and then briefly describe 
how to obtain the same quantity from the nonperturbative expression
for the Schwinger mechanism that includes
all order interactions with the background field.

\subsection{Lowest order perturbation}
We treat the interaction of the electron with the background 
${\mathcal L}_{\rm BG}$ as 
perturbation and compute an $S$-matrix element for the $e^+e^-$ pair creation from the vacuum, $S \equiv \braket{e^-(\mbox{\boldmath $p$},s)\, 
e^+(\mbox{\boldmath $p$}',s');{\rm  out} |{\rm  vac;in}} 
= \bra{e^-(\mbox{\boldmath $p$},s)\, e^+(\mbox{\boldmath $p$}',s'); 
{\rm in}} 
{\sf T}\exp[i\int d^4x {\mathcal L}_{\rm BG} ] 
\ket{\rm vac;in}$, in the lowest order perturbation theory.  The diagrammatic expression for the lowest-order contribution 
$S^{(1)}$ reads
\begin{align}
	S^{(1)} &\equiv \bra{ e^-(\mbox{\boldmath $p$},s)\, 
                              e^+(\mbox{\boldmath $p$}',s');{\rm in} } 
 i\int d^4x {\mathcal L}_{\rm BG} \ket{\rm vac;in}  \nonumber\\
&=  \parbox{12mm}{\includegraphics[width=24mm]{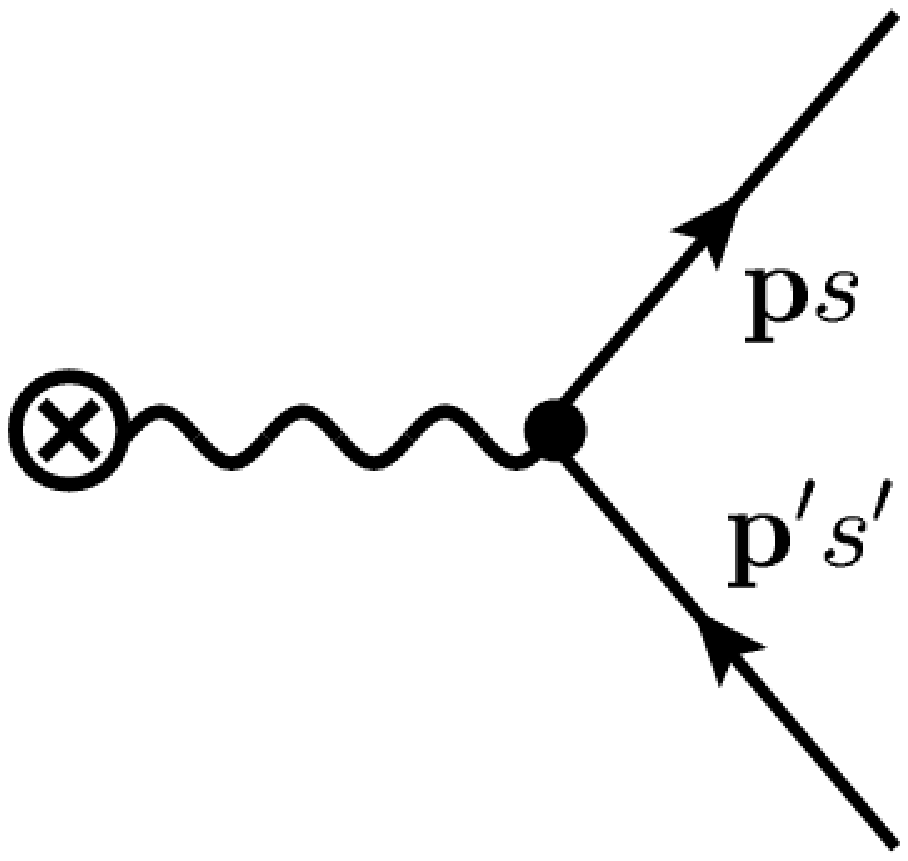}}\, .  
\end{align}
A straightforward calculation yields
\begin{align}
S^{(1)} 
	  &= -i \int d^4 x (e \bar{A}_{3}(x^0))
         \bra{e^-(\mbox{\boldmath $p$},s)\, 
              e^+(\mbox{\boldmath $p$}',s') {\rm ;in}} 
         \bar{\psi}(x) \gamma^3 \psi(x) \ket{\rm vac;in} \nonumber \\
	  &= i \left[
\int dx^0 \left(e \bar{A}_{3}(x^0)\right) {\rm e}^{2 i p_0 x^0 } \right]  
\big( \bar{u}(\mbox{\boldmath $p$}, s) \gamma^3  v(-\mbox{\boldmath $p$}, s')  
\big) \delta^3({\mbox{\boldmath $p$}}+{\mbox{\boldmath $p$}}')
\nonumber \\
	  &= \frac{ e \tilde{E}(2p_0)}{2p_0} \;
\big( 
\bar{u}(\mbox{\boldmath $p$}, s) \gamma^3  
v( -\mbox{\boldmath $p$}, s')  \big) \delta^3(\mbox{\boldmath $p$}+\mbox{\boldmath $p$}')
\, . \label{eq4}
\end{align} 
Note that we have expanded the unperturbed operator $\psi$ as 
\begin{align}
	\psi(x) =  \sum_s \int d^3\mbox{\boldmath $p$} 
\left[  u(\mbox{\boldmath $p$},s) 
\frac{{\rm e}^{-i(p_0x^0 - {\small \mbox{\boldmath $p \cdot x$}})}}
     {\sqrt{(2\pi)^3}}  a(\mbox{\boldmath $p$},s)  
    +  v(\mbox{\boldmath $p$},s) 
\frac{{\rm e}^{ i(p_0x^0 - {\small \mbox{\boldmath $p \cdot x$}})}}
     {\sqrt{(2\pi)^3}} b^{\dagger}(\mbox{\boldmath $p$},s)
   \right] 
\end{align}
to get the second expression. Here $p_0=\sqrt{m^2+\mbox{\boldmath $p$}^2}$,
the annihilation operators $a(\mbox{\boldmath $p$},s)$ for an electron 
and $b(\mbox{\boldmath $p$},s)$ for a positron satisfy the following 
anticommutation relations:
$
\big\{a(\mbox{\boldmath $p$},s),a^{\dagger}({\mbox{\boldmath  $p$}'},s')\big\} 
= \big\{b(\mbox{\boldmath $p$},s),b^{\dagger}({\mbox{\boldmath $p$}'},s')\big\} 
= \delta_{ss'}\delta^3(\mbox{\boldmath $p-p'$})\, ,\ 
	(\text{otherwise}) = 0\, ,
$
and the Dirac spinors $u(\mbox{\boldmath $p$},s), 
v(\mbox{\boldmath $p$},s)$ are normalized as
$
 u^{\dagger}(\mbox{\boldmath $p$},s) u(\mbox{\boldmath $p$},s')
 = v^{\dagger}(\mbox{\boldmath $p$},s) v(\mbox{\boldmath $p$},s') 
= \delta_{ss'}\, , \ 
 v^{\dagger}(\mbox{\boldmath $p$},s) u(\mbox{\boldmath $p$},s') = 0\, .  
$ 
In the last line of Eq.~(\ref{eq4}) we have introduced the Fourier transform of the background electric field $\tilde E(\omega)= 
\int dt E(t) \, {\rm e}^{i\omega t}$.

Now, we can compute the number density of electrons 
$d^3 N/d \mbox{\boldmath $p$}^3$ 
created from the vacuum in the lowest-order perturbation theory, 
\begin{align}
	\frac{1}{V} \frac{d^3 N}{d\mbox{\boldmath $p$}^3} 
 &= \frac{1}{V}\sum_{s'} \int d^3\mbox{\boldmath $p$}' \big|S^{(1)}\big|^2
= \frac{1}{(2\pi)^3}\left( 1 - \frac{p_z^2}{p_0^2} \right) 
\frac{ \big| e \tilde{E}(2 p_0)\big|^2}{4 p_0^2}    
\, ,  \label{eq15}
\end{align}
where integration over the positron momentum $\mbox{\boldmath $p$}'$ 
gives a volume factor $V=(2\pi)^3 \delta^3({\bf 0})$.  
The physical meaning of the formula (\ref{eq15}) is evident. 
For an electric field oscillating in time 
$E(x^0)=E_0 \cos \omega x^0$, $\tilde{E}(2p_0)$ is proportional to $\delta(2p_0+\omega) + \delta(2p_0-\omega)$. For the on-shell
electron energy $p_0>m$, the number of produced electrons 
vanishes if $|\omega|<2m$, which means that the pair creation does not
occur when the energy supplied by a single photon is below this threshold.
This is certainly true for a constant electric field $\omega\to 0$, 
no matter how strong the background electric field is 
(within the perturbation theory). 
For a general time-dependent background field
the number of 
produced electrons is nonvanishing even for a single photon 
as long as the background electric field 
has a nonzero Fourier spectrum $\tilde E(\omega)$
above the threshold $\omega\ge 2m$.  

The total number of produced electrons, $N$, is obtained after 
integration over \mbox{\boldmath $p$}: 
\begin{align}
	\frac{N}{V}
&=
\frac{1}{(4\pi)^2}\int_{2m}^{\infty} d\omega \; \sqrt{1-\frac{4m^2}{\omega^2} }
\frac{1}{3} \left ( 2+\frac{4m^2}{\omega^2} \right ) 
\big|e {\tilde E}(\omega)\big|^2
\; .
\label{eq20}
\end{align}
The $\omega$ integral is not possible in general unless 
we specify the background electric field $\tilde{E}$.

\subsection{Nonperturbative evaluation -- Schwinger mechanism}

Creation of $e^+e^-$ pairs from the vacuum is possible 
in the presence of an electric field as a nonperturbative process, which is characterized by 
the critical field strength $E_c=m^2/e$.  Schwinger \cite{sch51} computed formulas for the number of created pairs and for the 
vacuum persistent probability in case of the electric fields which is homogeneous in space and constant in time.  It is also known that we can equally 
formulate the case with temporal dependence. Here, we briefly 
explain such a case. Notice that this is a nonperturbative calculation
because we include the interaction of electrons or positrons with the 
electric fields up to infinite order.

We employ a formalism based on the canonical quantization in 
the presence of external fields.  Namely, we expand the field 
operator $\psi$ by the exact mode functions 
of an electron ${}^{}_{+}\! \psi^{\rm as}_{\mbox{\boldmath {\small $p$}}s}$ 
(as\! =\! in/out) and a positron 
${}^{}_{-}\! \psi^{\rm as}_{\mbox{\boldmath {\small $p$}}s}$ 
(as\! =\! in/out) under the given background field $\bar{A}^\mu$ as
\begin{align}
&\psi (x) = \sum_s \int d^3\mbox{\boldmath $p$} 
\left[  {}^{}_+\! \psi^{\rm as}_{\mbox{\boldmath {\small $p$}}s}(x) 
a^{\rm as}(\mbox{\boldmath $p$},s) + 
{}^{}_{-}\!\psi^{\rm as}_{\mbox{\boldmath {\small $p$}}s}(x) 
b^{{\rm as}\dagger}(-\mbox{\boldmath $p$},s) \right]\, , \\
&\left[ i\Slash{\partial} - e\bar{\Slash{A}} - m\right] 
{}^{}_{\pm}\! \psi^{\rm as}_{\mbox{\boldmath {\small $p$}}s}(x)=0\quad  {\rm (as=in/out)}
\; . 
\label{eq23}
\end{align}
Noting that any linear combinations of 
$_\pm\! \psi^{\rm as}_{\mbox{\boldmath {\small $p$}}s}(x)$ satisfy
the equation of motion (\ref{eq23}), we identify the electron (positron)
mode as the positive (negative) frequency mode 
in the asymptotic in- and out-region, respectively.
The important point is that the mode
function of the in-state 
${}^{}_{\pm}\! \psi^{\rm in}_{\mbox{\boldmath {\small $p$}}s}$ and that of the out-state 
${}^{}_{\pm}\! \psi^{\rm out}_{\mbox{\boldmath {\small $p$}}s}$ do not coincide with each other in the presence of the background field 
and the electron (positron) mode function of the in-state 
 ${}^{}_{+}\! \psi^{\rm in}_{\mbox{\boldmath {\small $p$}}s}$ 
(${}^{}_{-}\! \psi^{\rm in}_{\mbox{\boldmath {\small $p$}}s}$) 
becomes a linear combination of the mode functions of the out-state: 
\begin{align}
  \begin{pmatrix} 
  ^{}_+\!\psi^{\rm in}_{\mbox{\boldmath {\small $p$}}s} \\ 
  ^{}_-\!\psi^{\rm in}_{\mbox{\boldmath {\small $p$}}s}
  \end{pmatrix}
&= 
  \begin{pmatrix} 
   \alpha^{}_{\mbox{\boldmath {\small $p$}}} 
& -\beta^{*}_{\mbox{\boldmath {\small $p$}}} \\  
   \beta^{}_{\mbox{\boldmath {\small $p$}}}  
&  \alpha^{*}_{\mbox{\boldmath {\small $p$}}}
  \end{pmatrix}   
  \begin{pmatrix} 
  ^{}_+\!\psi^{\rm out}_{{\mbox{\boldmath {\small $p$}}} s} \\ 
  ^{}_-\!\psi^{\rm out}_{{\mbox{\boldmath {\small $p$}}} s} 
  \end{pmatrix}.  
\label{2bog}
\end{align}
The Bogoliubov coefficients, $\alpha_{\mbox{\boldmath {\small $p$}}}$ 
and $\beta_{{\mbox{\boldmath {\small $p$}}}}$,
satisfy the relation $|\alpha_{\mbox{\boldmath {\small $p$}}}|^2 
+ |\beta_{{\mbox{\boldmath {\small $p$}}}} |^2 =1$.
One may simply understand this fact (\ref{2bog})
in analogy with an one-dimensional barrier scattering problem where we always have a mixture of in-coming wave and its reflection on one side of the barrier, but the other side consists of out-going wave only.  This difference between in- and out-state mode functions 
results in the difference between 
the in- and out-state annihilation operators, 
$a^{\rm as}({\mbox{\boldmath {$p$}}},s)$ and  
$b^{\rm as}({\mbox{\boldmath {$p$}}},s)$.  
From the orthonormality of the mode functions, one can obtain
\begin{align}
a^{\rm out}({\mbox{\boldmath $p$}},s) 
&= \int d^3{\mbox{\boldmath $x$}}\, 
   \big(
     {}^{}_+\! \psi^{{\rm out}}_{{\mbox{\boldmath {\small $p$}}}s}(x)
\big)^\dagger \psi(x) 
= \alpha^{}_{{\mbox{\boldmath {\small $p$}}}} 
  a^{\rm in}({\mbox{\boldmath $p$}},s)  
+ \beta^{}_{{\mbox{\boldmath {\small $p$}}}} 
  b^{{\rm in}\dagger}(-{\mbox{\boldmath $p$}},s)\, , \label{eq24}\\
 b^{\rm out}(-{\mbox{\boldmath $p$}},s) &= 
\int d^3{\mbox{\boldmath $x$}}\, 
\big( {}^{}_-\! \psi^{{\rm out}}_{{\mbox{\boldmath {\small $p$}}}s}(x)
\big)^{\dagger} \psi(x) 
= -\beta^*_{{\mbox{\boldmath {\small $p$}}}} 
   a^{\rm in}({\mbox{\boldmath $p$}},s) 
+ \alpha^*_{{\mbox{\boldmath {\small $p$}}}} 
   b^{{\rm in}\dagger}(-{\mbox{\boldmath $p$}},s).  \label{eq25}
\end{align}
Using Eqs.~(\ref{eq24}) and (\ref{eq25}), we can construct a nonperturbative formula to compute the number of electrons created via the Schwinger mechanism: 
\begin{align}
\frac{1}{V}\frac{d^3 N}{d\mbox{\boldmath $p$}^3} 
= \frac{1}{V}\bra{{\rm vac;in}   }  
a^{{\rm out}\dagger}(\mbox{\boldmath $p$},s) 
a^{\rm out}(\mbox{\boldmath $p$},s)  
\ket{{\rm vac;in}   } 
= \frac{1}{(2\pi)^3}| \beta_{{\mbox{\boldmath {\small $p$}}}}|^2.  
\label{eq26} 
\end{align}
Thus, the problem now is reduced to computing 
the coefficient $\beta^{}_{\mbox{\boldmath {\small $p$}}}$ or solving the 
Dirac equation (\ref{eq23}).  
Notice that we have not specified the time dependence of 
the background field $\bar{A}^\mu$ so far and thus Eq.~(\ref{eq26}) is 
a general formula.  
However, since there are only a few cases where analytic 
solutions for the Dirac equation (\ref{eq23}) are available, we are 
usually forced to use numerical methods to evaluate Eq.~(\ref{eq26}).  When the background electric field is constant and homogeneous,
one can easily solve the Dirac equation (\ref{eq23}) \cite{NikishovJETP30.666}. The total 
number of produced electrons per unit volume and time is given by 
$N/(TV)\propto |eE|^2 \exp \{-\pi m^2/|eE|\}$, which clearly shows 
the nonperturbative nature of the formula (\ref{eq26}).

\section{PULSED ELECTRIC FIELD: SAUTER-TYPE BACKGROUND FIELD} \label{sec3}

In this section, we consider a special case where the background 
electric field is applied as a pulse in time. In particular, 
we work with a Sauter-type pulse field\footnote[4]{Originally, Sauter \cite{sau32}
considered the Dirac equation in an inhomogeneous potential 
$V(z)=V_0 {\rm sech}^2 (z/d)$. Since the problem is essentially 
reduced to solving a differential equation in one dimension, 
we can equally solve the potential with the similar functional 
dependence on time. } with height $E_0$ and width $\tau$: 
\begin{align}
\bar{A}_3(t) = E_0 \tau \tanh(t/\tau)\ \ {\rm or}\ \ 
E(t) = E_0\, {\rm sech}^2(t/\tau).   \label{eq21}
\end{align}
As mentioned in the Introduction, the advantage of the Sauter-type 
background field is that the analytic solution for the Dirac equation is known so that
we can explicitly compute the particle number 
created via the Schwinger mechanism \cite{NaroNiki,SkokovLevai09,Hebenstreit}. 
Thus, by comparing this nonperturbative result with that of our 
perturbative computation obtained in the Sauter-type background 
field, we can discuss which picture, 
perturbative or nonperturbative, is appropriate for studying the pair production in a strong field with finite duration.

\subsection{Perturbative result}

We substitute the Sauter-type background field (\ref{eq21}) into 
Eqs.~(\ref{eq15}) and (\ref{eq20}) to get the 
electron number density $d^3 N /d\mbox{\boldmath $p$}^3$ and the 
total electron number $N$, respectively, in the lowest-order perturbation theory.  By using
\begin{align}
\tilde{E}(\omega)=i\omega \tilde{A}_3(\omega)
=  \frac{i\pi E_0 \tau^2 \omega}{\sinh \frac{\pi \tau \omega}{2}}\, ,
\label{spect}
\end{align}
we find that the electron number density $d^3 N /d\mbox{\boldmath $p$}^3$ 
is given by 
\begin{align}
\frac{1}{V}\frac{d^3 N}{d\mbox{\boldmath $p$}^3} 
= \frac{1}{(2\pi)^3}\left( 1-\frac{p_z^2}{p_0^2} \right) \left|  
\frac{e E_0}{p^2_0}\right|^2   \frac{(\pi p_0 \tau)^4}{\pi^2 \left|{\rm sinh}[\pi p_0 \tau]\right|^2}
\, . \label{eq233}
\end{align}
We obtain a nonvanishing result because the Fourier spectrum of 
the Sauter-type field $\tilde{E}(\omega)$ is nonzero 
at any value of $\omega$, in particular in the region $\omega\ge 2m$.  Also, the total electron number $N$ 
is given by
\begin{align}
\frac{N}{V}
 = m^3 \left| \frac{eE_0 }{m^2}   \right|^2 f(\pi m \tau)\, .  
\label{eq244}
\end{align}
Here we have introduced the function $f$ by
\begin{align}
f(x) \equiv \frac{x^4}{2\pi^4} \int_{1}^{\infty} d\omega \omega^2 
\sqrt{1-\frac{1}{\omega^2}}\frac{1}{3}\left( 2 + \frac{1}{\omega^2} \right)
\frac{1}{\left| \sinh(\omega x) \right|^2}\, , \label{eq222}
\end{align}
which behaves asymptotically as (see the Appendix)
\begin{align}
f(x) \sim \left\{  \begin{array}{ll} \displaystyle \frac{x}{18\pi^2} 
& (x\lesssim 1) \\ & \\
\displaystyle \frac{ x^{5/2}}{2 \pi^{7/2}}\left( 1 + \frac{7}{16}\frac{1}{x}  \right) {\rm e}^{-2x} & (x\gtrsim 1).  \end{array} \right. \label{eq188}
\end{align}

\begin{figure}[t]
  \begin{center}
   \includegraphics[angle=-90,width=0.6\textwidth]{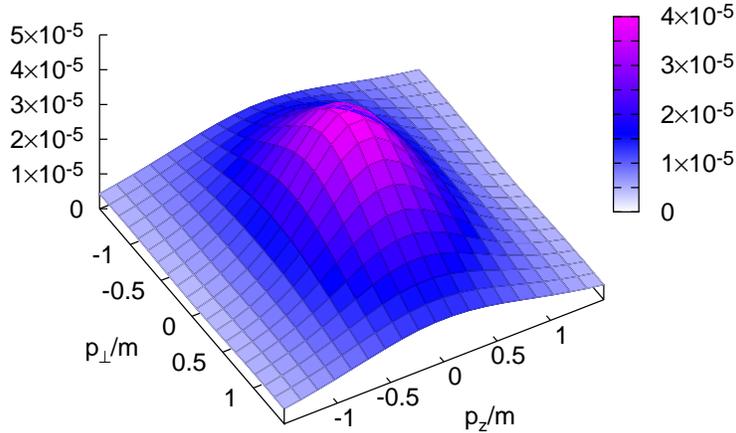}
  \end{center}
  \caption{(color online).  The momentum ($\mbox{\boldmath $p$}$) dependence of the number density of electrons $(1/V) d^3 N/d\mbox{\boldmath $p$}^3$ (\ref{eq233}).  Parameters are set to $|eE_0|/m^2 = 10$ and $m\tau =0.01$.}  
  \label{fig1}
\end{figure}

\begin{figure}[t]
  \begin{center}
   \includegraphics[angle=-90,width=0.6\textwidth]{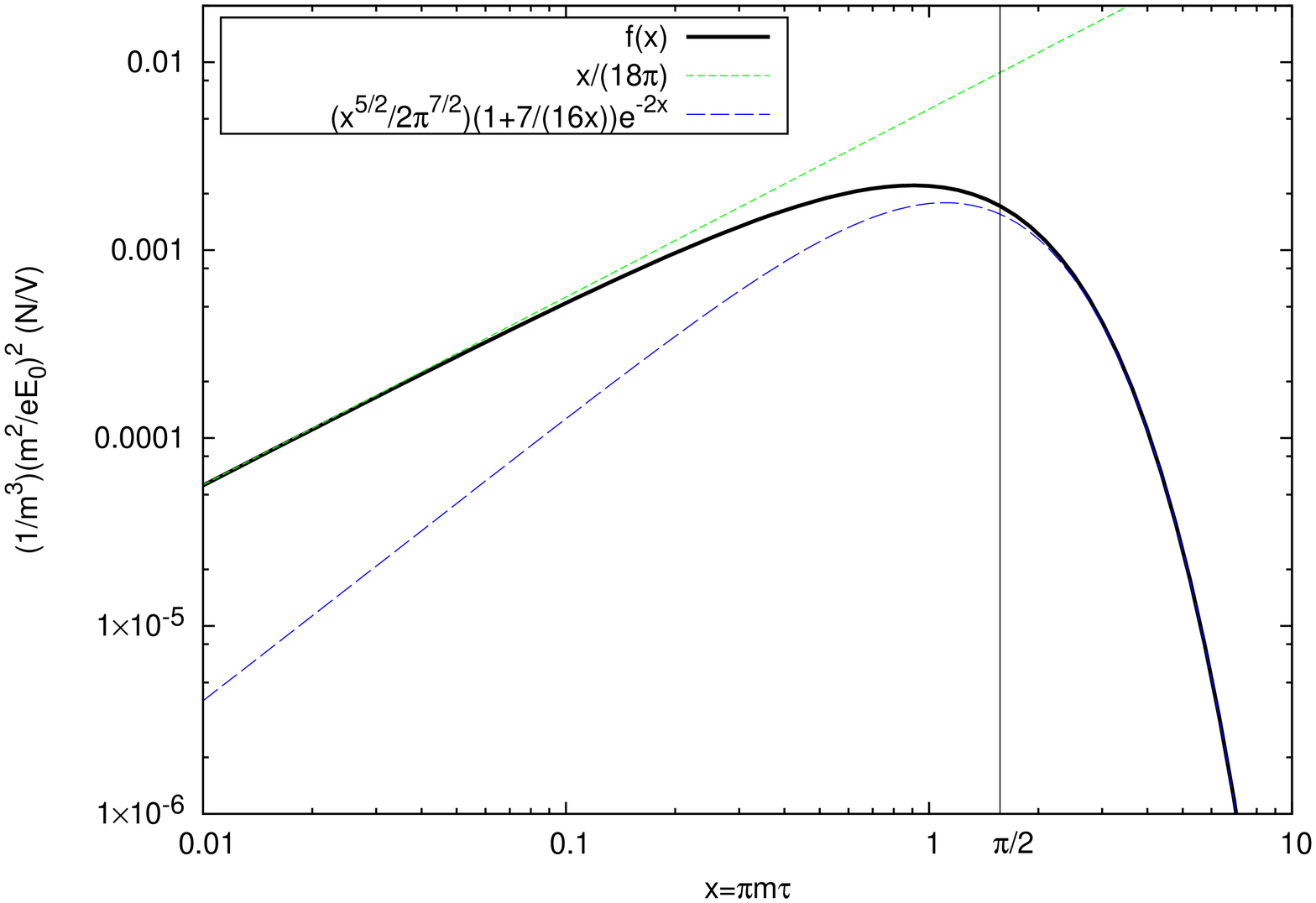}
  \end{center}
  \caption{(color online).  The total number of electrons $N/V$ as a function of $\pi m \tau$
in the lowest-order perturbation theory (\ref{eq244}) (solid line). 
Two asymptotic forms (\ref{eq188}) are shown in dashed and dotted lines.
The vertical black line indicates the point $1/\tau = 2m$.}
  \label{fig2}
\end{figure}

Figure~\ref{fig1} shows the momentum ($\mbox{\boldmath $p$}$) 
dependence of the 
electron number density $(1/V) d^3 N/d\mbox{\boldmath $p$}^3$ (\ref{eq233}) 
for $|eE_0|/m^2 = 10$ and $m\tau =0.01$.  
We see that the peak is located at $\mbox{\boldmath $p$}={\bf 0}$, 
which reflects the fact that the energy threshold 
for creating one $e^+ e^-$ pair ${\mathcal E}_{\rm thres} = 2p_0$ 
takes its minimum ${\mathcal E}_{\rm thres} = 2m$ at $\mbox{\boldmath $p$}={\bf 0}$.  
We also find that the distribution decays exponentially for large $p_\perp$.  
Actually, one can check this by taking the limit of $p_\perp \gg p_z, m$: 
\begin{align}
	\frac{1}{V}\frac{d^3 N}{d\mbox{\boldmath $p$}^3} \overset{p_\perp \gg p_z, m}{\longrightarrow} \frac{1}{2\pi} \left| \frac{eE_0}{m^2}  \right|^2 (m\tau)^4 {\rm e}^{-2\pi |\mbox{\boldmath {\small $p$}}_{\perp}| \tau}.
\end{align}

\begin{figure}[t]
  \begin{center}
   \includegraphics[angle=-90,width=0.45\textwidth]{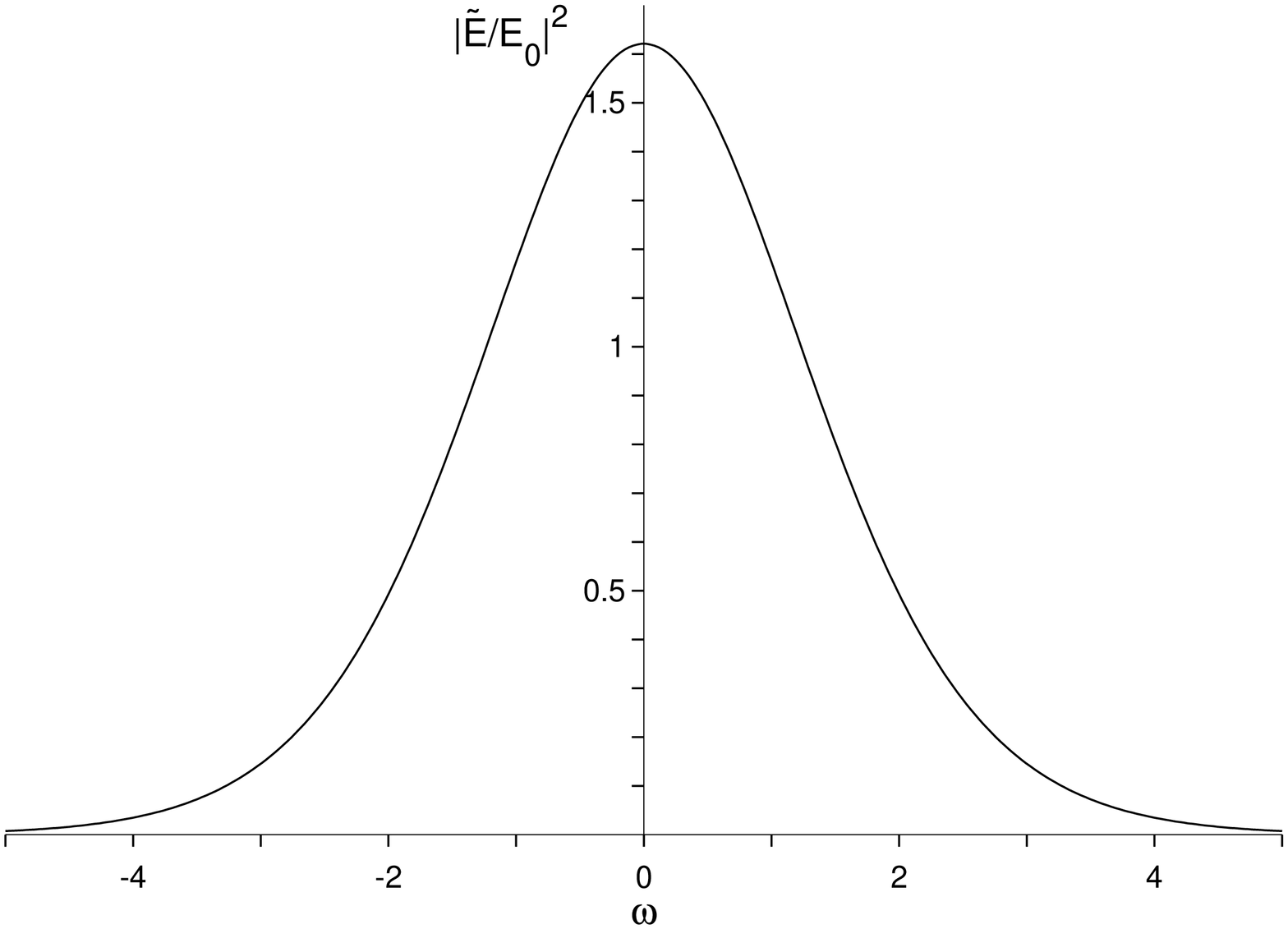}
  \hspace{8mm}
   \includegraphics[angle=-90,width=0.45\textwidth]{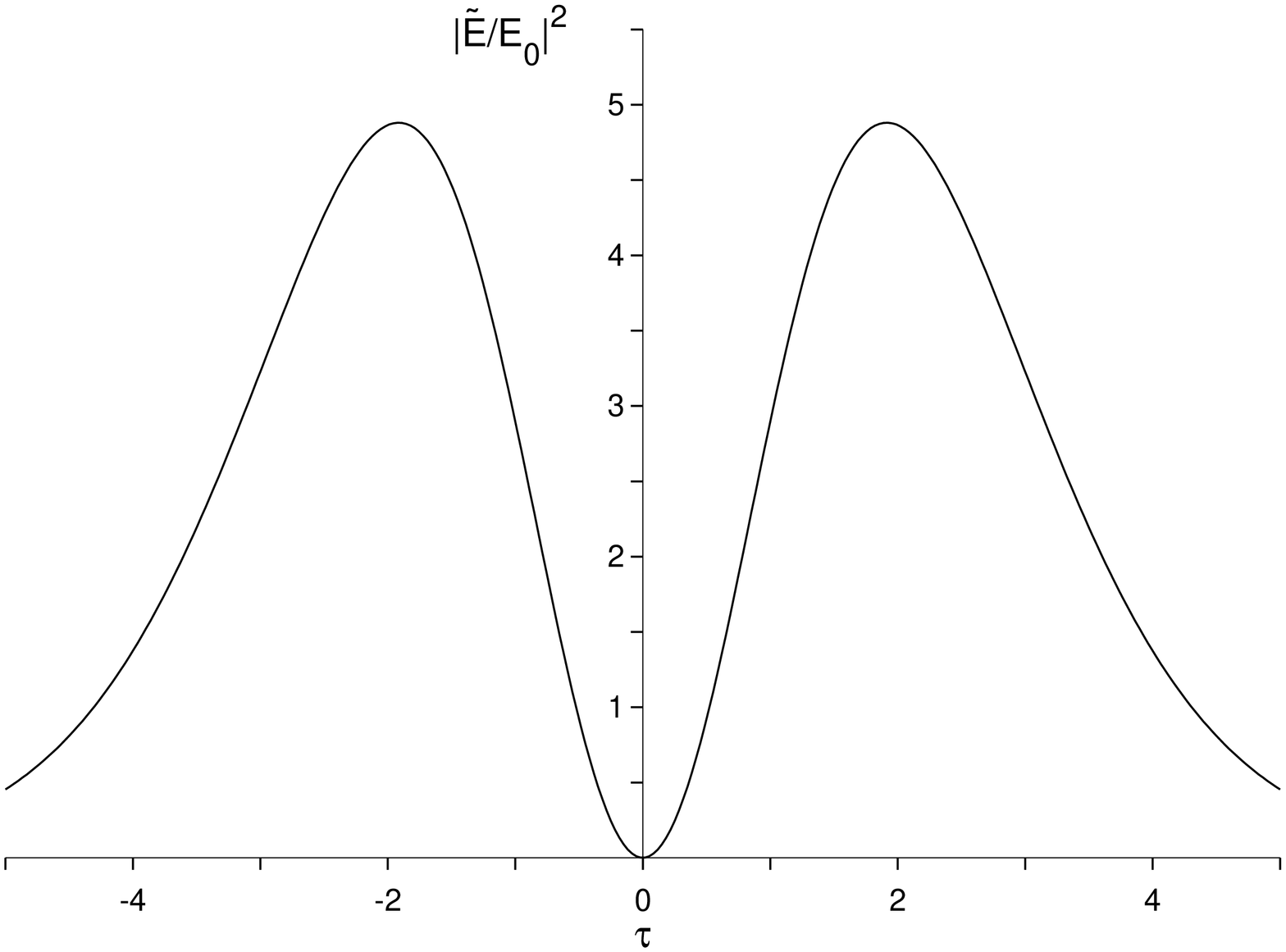}
  \end{center}
  \caption{Spectrum $|\tilde E(\omega)|^2$ of the Sauter-type electric field: 
  $\omega$ dependence for $\pi\tau /2=1$ (left) and 
  $\tau$ dependence for $\pi \omega /2=1$ (right).  }
  \label{spectrum}
\end{figure}

Figure~\ref{fig2} shows the $\tau$ dependence of the total electron 
number $N/V$.  As is seen in Fig.~\ref{fig2}, $N$ increases monotonically 
for small $m\tau$ while it decreases exponentially for large $m \tau$.  
This tendency can be roughly understood as follows: Since the threshold 
energy of the pair creation is ${\mathcal E}_{\rm thres} \sim 2m$, 
the background field must supply energy $\Omega$ larger than $2m$ 
for the pair creation to occur.  In our computation based on the 
lowest-order perturbation theory, such energy $\Omega$ is supplied 
by a single (virtual) photon from the background electric field $E$.  Since the typical energy $\omega$ of a photon 
which forms the Sauter-type background field $E$ is $\omega \sim 1/\tau$ 
(see Eq.~(\ref{spect}) and the left panel of Fig.~\ref{spectrum}), 
we find $\Omega\sim$ (number of photons) $\times$ (typical photon energy) 
$ \sim 1\times \omega \sim 1/\tau$.  
Thus, $\Omega \gtrsim 2m$ i.e., $m\tau \lesssim 1/2$ 
is required for the pair creation in the lowest order perturbation theory. 
The upper limit $m\tau=1/2$ is shown as a vertical line in 
Fig.~\ref{fig2}. Pair creation from a single photon occurs when the 
pulse duration $\tau$ is short enough. 
On the other hand, as shown in the right panel of Fig.~\ref{spectrum},
the strength of the Fourier component 
$\tilde E(\omega)$ decreases with decreasing $\tau$. This essentially 
explains the decrease of electron number density for $\tau\to 0$ 
as seen in Fig.~\ref{fig2}.

\subsection{Nonperturbative result}

One can obtain analytic solutions to the Dirac equation in the 
presence of the Sauter-type background field \cite{sau32,NaroNiki,Hebenstreit} 
(see also Ref.~\cite{SkokovLevai09} for discussion in $SU(2)$ case), 
which enables us to compute the number of produced electrons (\ref{eq26}). 
After some calculations, one finds
\begin{align}
\frac{(2\pi)^3}{V} \frac{d^3 N}{d\mbox{\boldmath $p$}^3} &=   
\frac{
\sinh\left[ \frac{\pi\tau}{2} 
            \left( 2eE_0 \tau  +p_{0}^{(-)} - p_{0}^{(+)}
            \right)  \right] 
\sinh\left[ \frac{\pi\tau}{2} 
            \left( 2eE_0 \tau  -p_{0}^{(-)} + p_{0}^{(+)}
            \right)  \right] 
}
{
\sinh\left[ \pi \tau p_{0}^{(-)} \right] 
\sinh\left[ \pi \tau p_{0}^{(+)} \right]
}
\, ,  \label{eqa16}
\end{align}
where 
$p_{0}^{(\pm)} \equiv \sqrt{m^2 + \mbox{\boldmath $p$}_\perp^2 + (p_z \pm eE_0 \tau)^2}$
are the energy of the electron (positron) with the transverse 
momentum $\mbox{\boldmath $p$}_\perp$ and the {\it canonical} longitudinal momentum $p_z$.
Note that the corresponding electron mode originally has 
the {\it kinetic} longitudinal momentum $p_z - eE_0 \tau$ in the infinite past and $p_z + eE_0 \tau$ in the infinite future.  We stress that this result is clearly nonperturbative with respect 
to the coupling constant $e$, while the lowest order perturbation gave 
the result proportional to $e^2$ [see Eq.~(\ref{eq233})].  
The total number of produced electrons, $N/V$, 
is obtained after integration
over the momentum $\mbox{\boldmath $p$}$.

\begin{figure}[t]
 \begin{center}
  \includegraphics[angle=-90,width=0.47\textwidth]{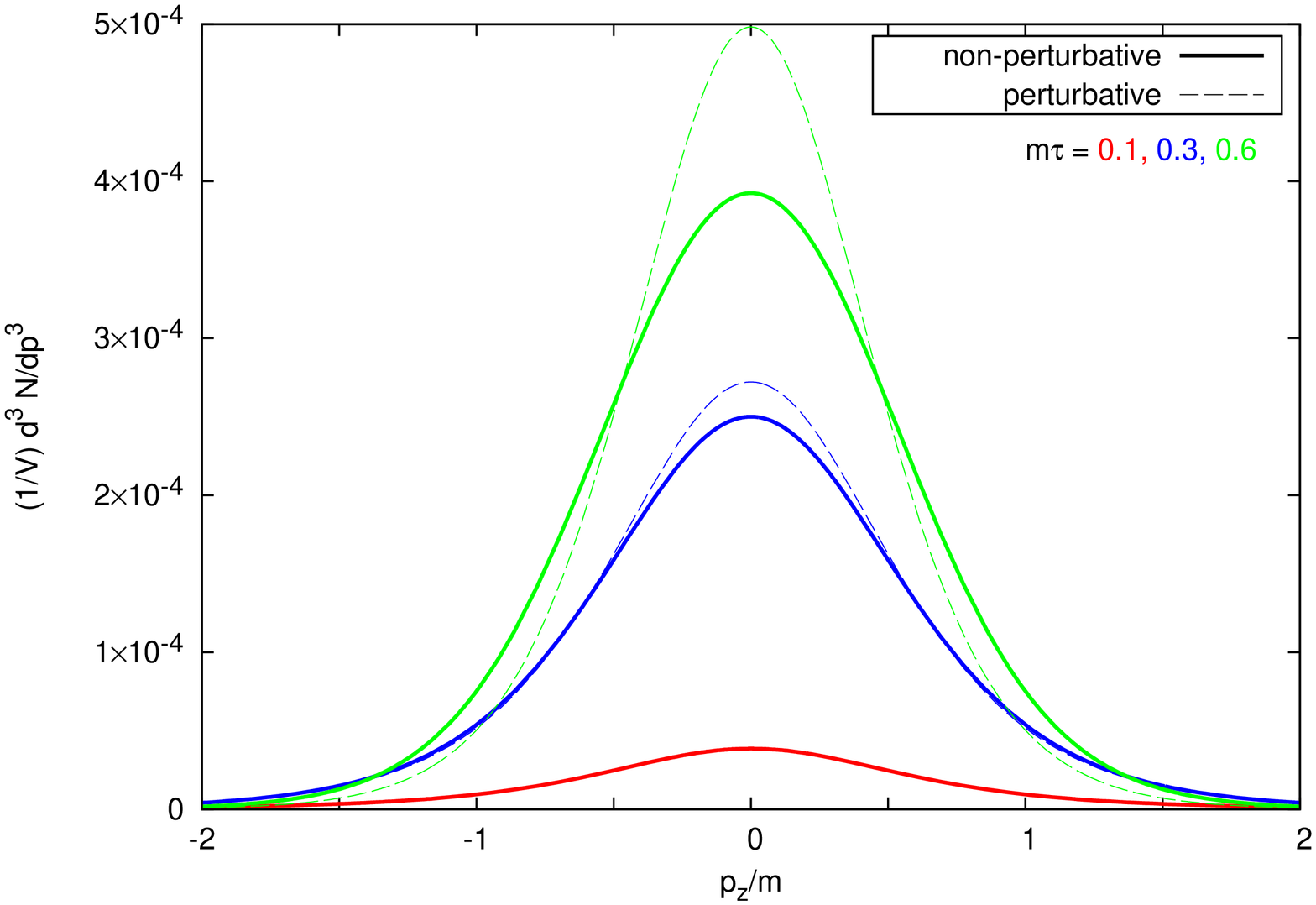}
\hfil
  \includegraphics[angle=-90,width=0.47\textwidth]{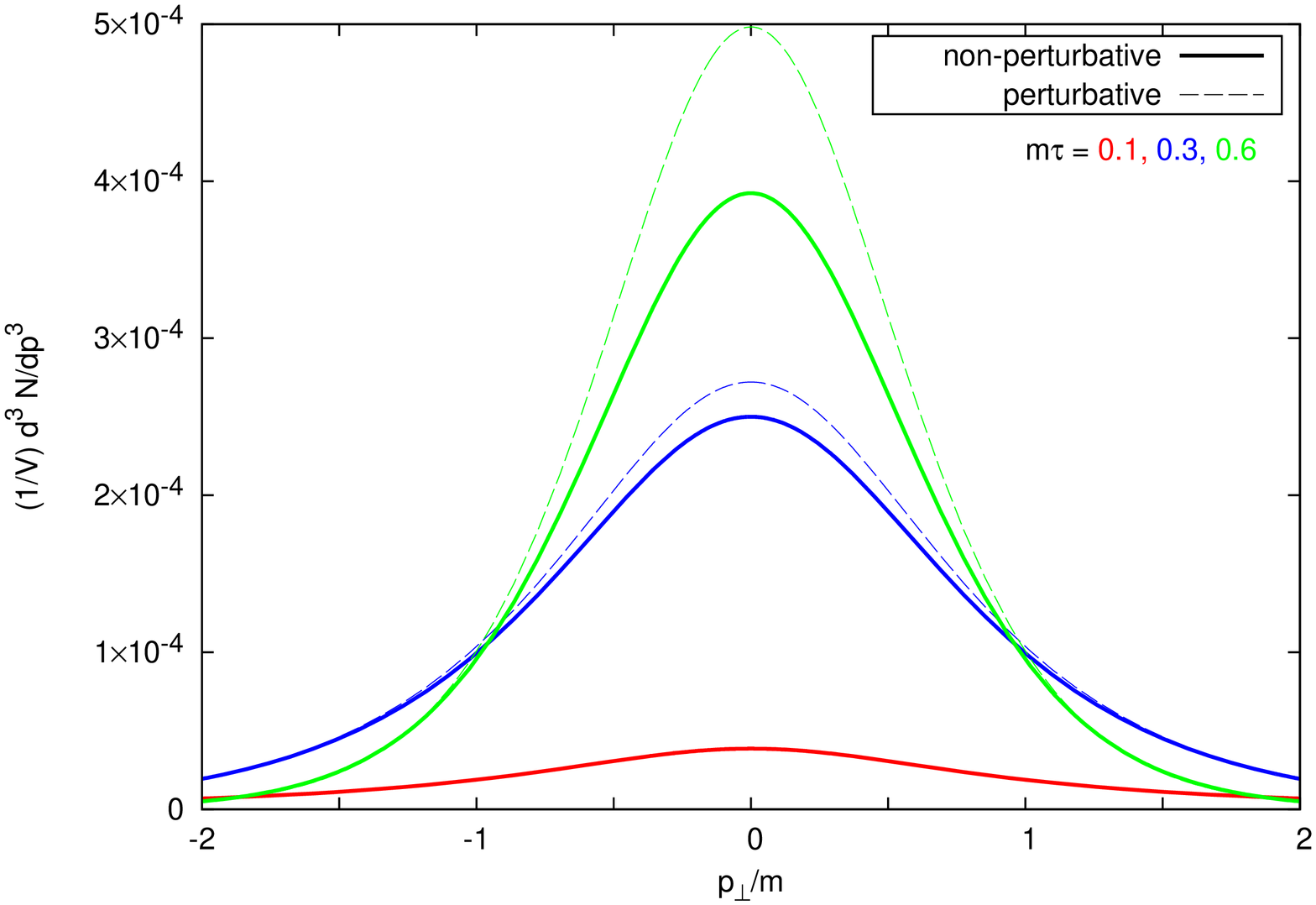}
\caption{(color online).  Comparison of the number density of 
electrons $d^3 N/d\mbox{\boldmath $p$}^3$.
 Solid lines represent the nonperturbative result (\ref{eqa16}) and dashed lines represent the perturbative result (\ref{eq233}) at $|eE_0|/m^2=1$.  
Left: $p_z$ dependence at $p_\perp/m=0$ with various duration $m \tau $.  Right: $p_\perp$ dependence at $p_z/m=0$ with various duration $m \tau$. }
  \label{Fig3}
 \end{center}
\end{figure}

\begin{figure}[t]
  \begin{center}
   \includegraphics[angle=-90,width=0.47\textwidth]{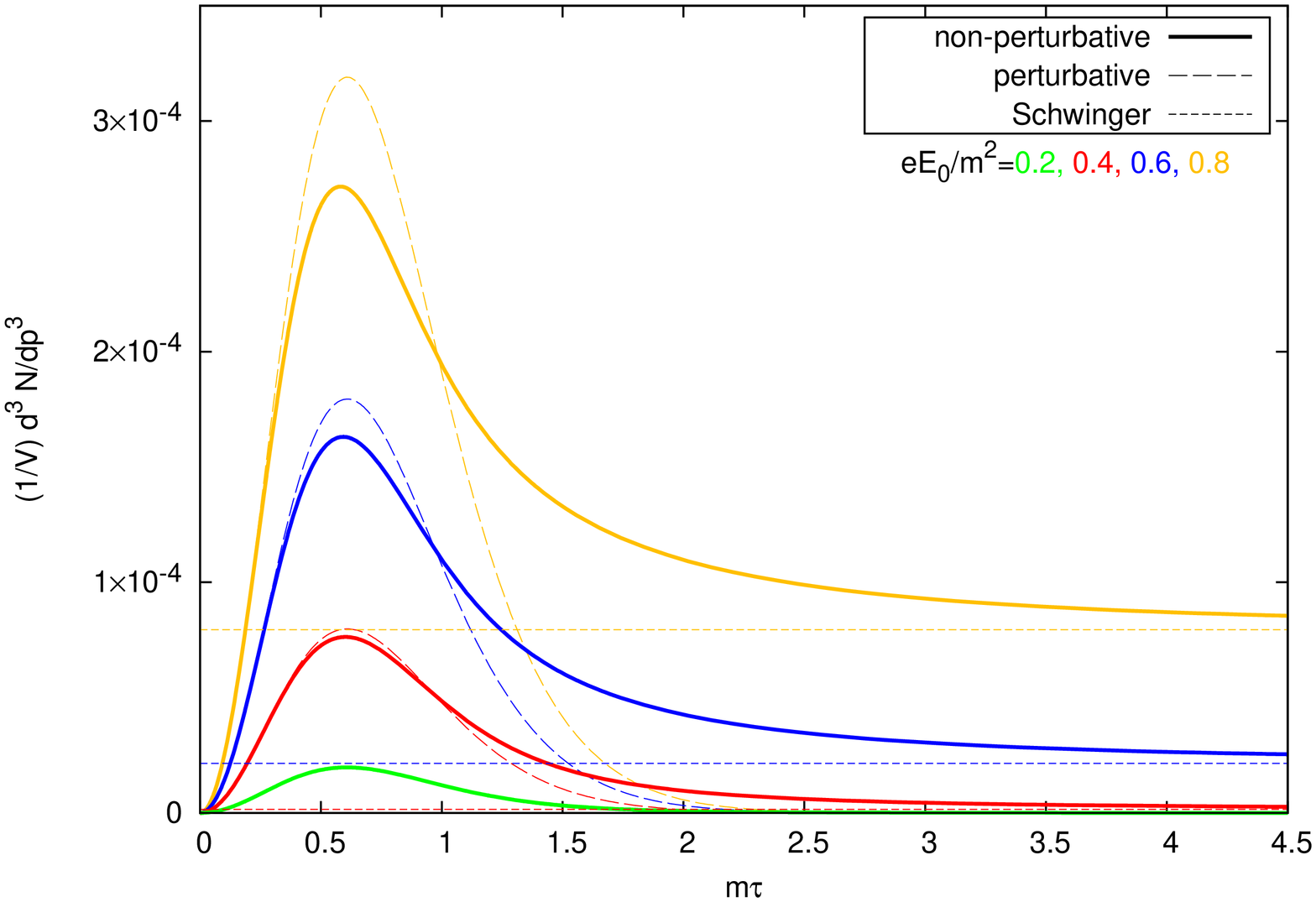}
  \hfil
   \includegraphics[angle=-90,width=0.47\textwidth]{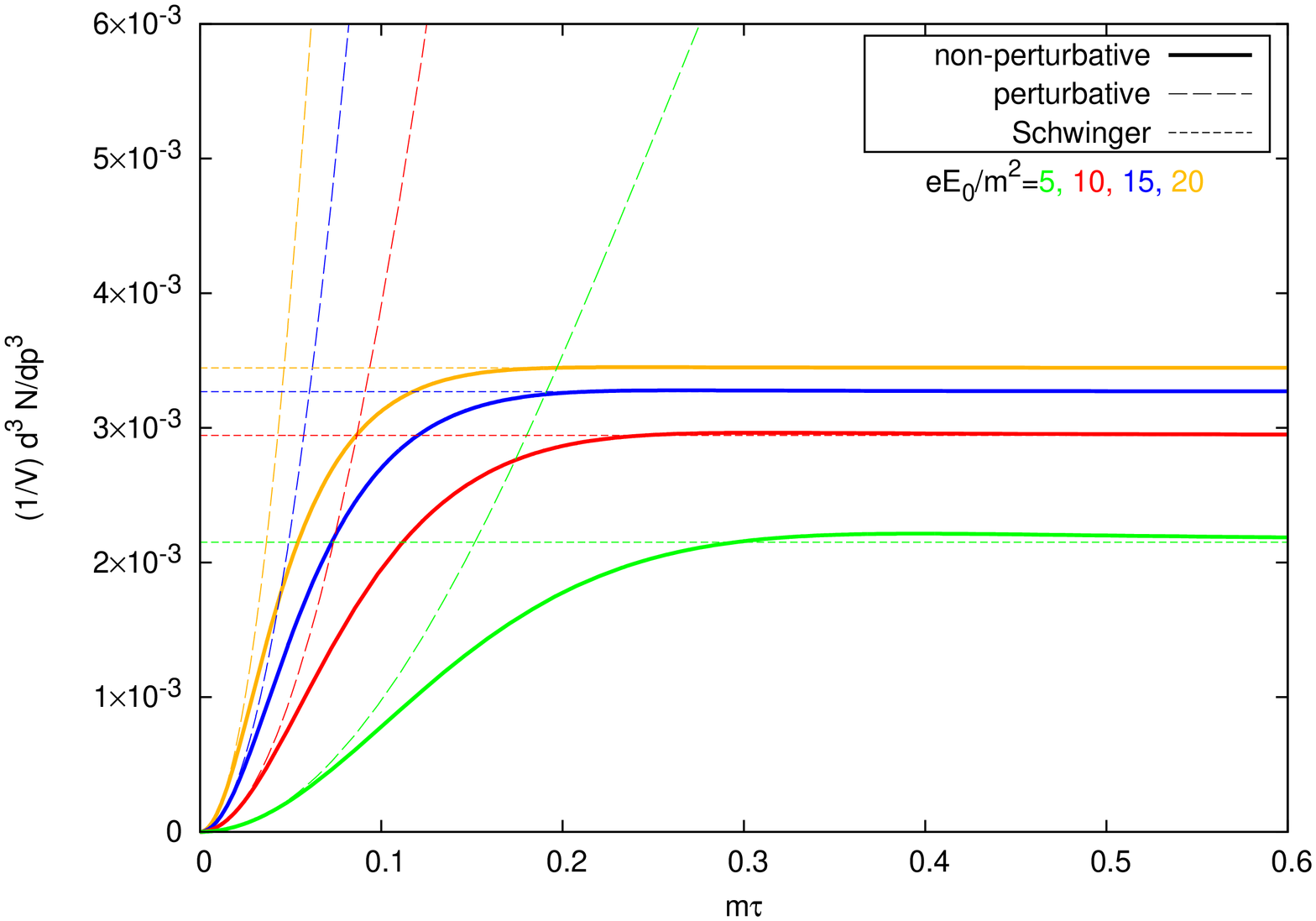}
	\caption{(color online).  Comparison of $\tau$ dependence of the number density of electrons 
$d^3 N/d\mbox{\boldmath $p$}^3$ 
at $p_z/m=p_\perp/m=0$ for the subcritical field strength $|eE_0|/m^2=0.2, 0.4, 0.6$ and 0.8 (left),
and for supercritical field strength $|eE_0|/m^2=5, 10, 15$ and 20 (right).  
Solid lines represent the nonperturbative result 
(\ref{eqa16}) and 
dashed lines represent the perturbative result (\ref{eq233}). 
The horizontal lines indicate Schwinger's result 
$(1/V)d^3 N/d\mbox{\boldmath $p$}^3 \big|_{\mbox{\boldmath {\small $p$}}
={\bf 0}}= \exp\{ -\pi m^2/|eE_0| \}/(2\pi)^3$ 
obtained in a constant electric field background.
 }
  \label{Fig3_taudep}
 \end{center}
\end{figure}

\begin{figure}[t]
  \begin{center}
   \includegraphics[angle=-90,width=0.47\textwidth]{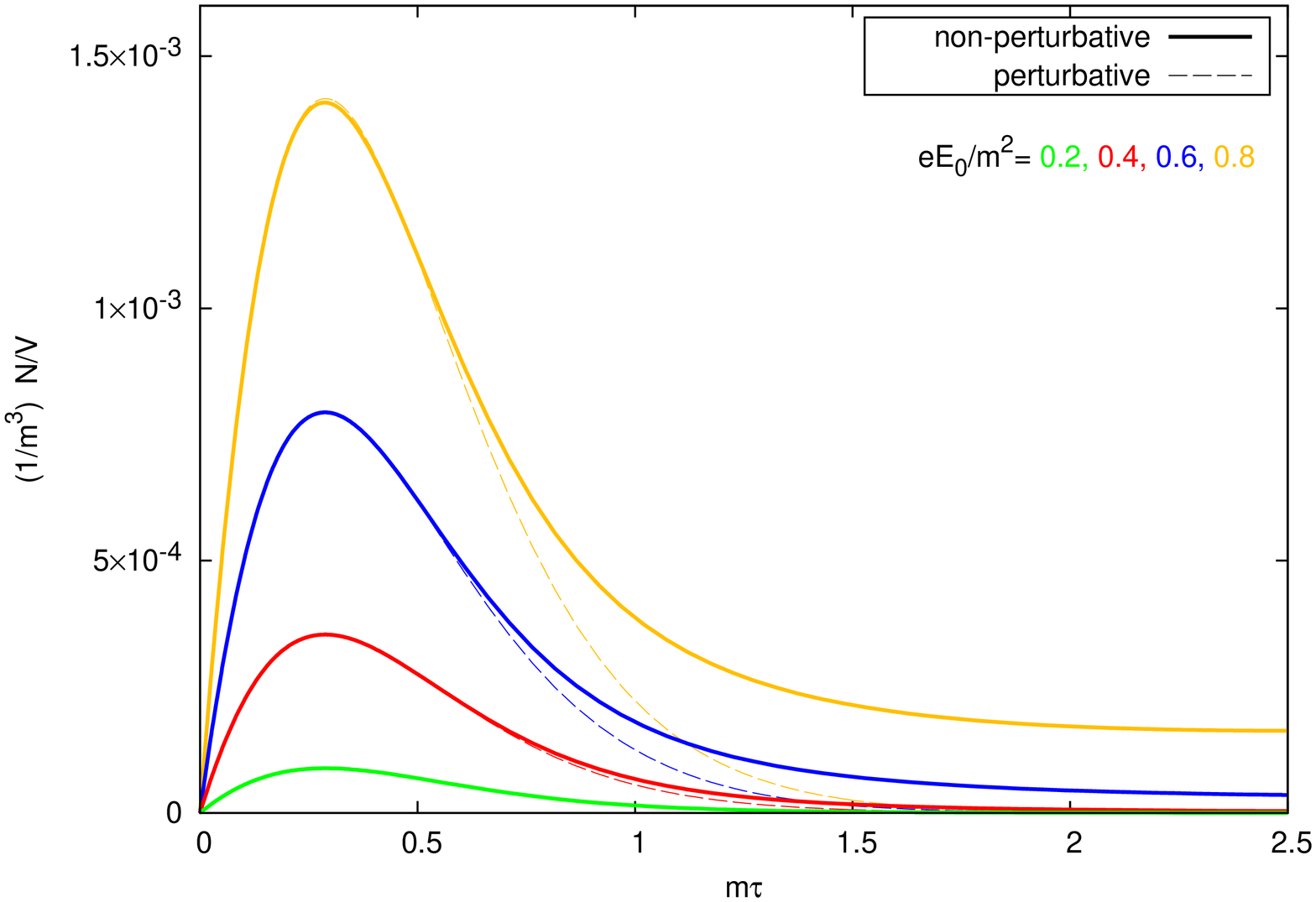}
  \hfil
   \includegraphics[angle=-90,width=0.47\textwidth]{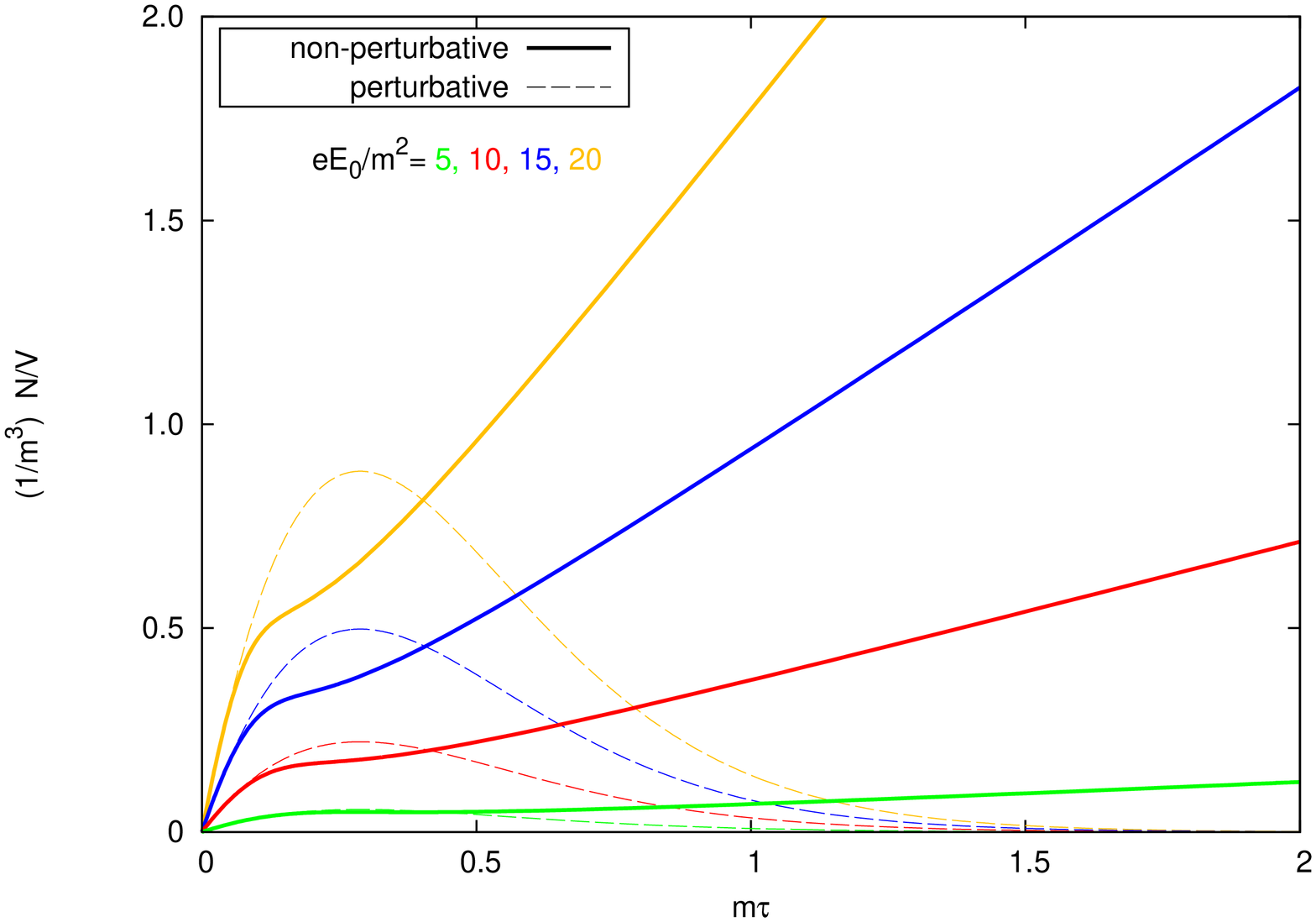}
	\caption{(color online).  Comparison of the total number of electrons $N$ as a function of duration $m\tau$ for subcritical field strength $|eE_0|/m^2=0.2, 0.4, 0.6$ and 0.8 (left), and 
for supercritical field strength $|eE_0|/m^2=5, 10, 15$ and 20 (right).
Dashed lines represent the perturbative result (\ref{eq244}) and
solid lines represent the nonperturbative result obtained by integrating (\ref{eqa16}) over $\mbox{\boldmath $p$}$.}
  \label{fig4}
 \end{center}
\end{figure}

\subsection{Comparison of the perturbative and nonperturbative results}
We compare the nonperturbative result (\ref{eqa16}) 
with the perturbative one (\ref{eq233}) and (\ref{eq244}) 
in Figs.~\ref{Fig3}, \ref{Fig3_taudep}, and \ref{fig4}. 
Figure~\ref{Fig3} shows the comparison of 
the momentum ($p_z$, $p_\perp$) dependence of 
the number density of electrons $d^3 N/d\mbox{\boldmath $p$}^3$. 
The peak strength of the field is taken as $|eE_0|/m^2=1$. 
Three lines are for different values of $\tau$.  We have shown only 
relatively short pulse cases: $m\tau=0.1,\, 0.3,$ and 0.6.
We immediately observe that the nonperturbative result (\ref{eqa16}) 
and the perturbative result (\ref{eq233}) coincide with each other 
for the {\it short} pulse. 
The deviation becomes larger as $\tau$ increases, which can be 
explicitly seen in Fig.~\ref{Fig3_taudep}.  
There, $\tau$ dependence is shown for different values of the peak 
strength: The left panel is for the subcritical\footnote{We tentatively 
use the words ``supercritical" and ``subcritical" for the cases 
$|eE_0|/m^2>1 $ and $|eE_0|/m^2<1$, respectively, but precisely speaking,
the condition $|eE_0|/m^2=1$ (valid for a constant electric field) 
does not play the same role for finite pulses.} 
field strength $|eE_0|/m^2 = 0.2,\, 0.4,\, 0.6$ and 0.8 and the right panel 
is for the supercritical field strength $|eE_0|/m^2 = 5,\, 10,\, 15$ and 20.  
We again observe the agreement of the two results 
for {\it short} pulses $m\tau\ll 1$ no matter how large the field strength 
$|eE_0| / m^2$ is.  However, the size of the agreement region in $\tau$ 
heavily depends on the field strength $|eE_0|/m^2$.  For subcritical field 
strength $|eE_0|/m^2 \lesssim 1$, perturbative result dominates the 
nonperturbative result even when the pulse is not {\it very short} $m\tau \sim 1$.  On the other hand, for supercritical field 
strength $|eE_0|/m^2 \gtrsim 1$, perturbative description is applicable 
only for {\it very short} pulse region $m\tau \ll 1$.  
We will clarify the reason for this behavior in the later discussion.  
The important point here is that 
for any field strength $|eE_0| / m^2$ there surely exists a region 
({\it short} pulse region) where pair creation can be understood as 
a purely perturbative phenomenon. 
We can also observe that there is a clear deviation between 
the two in the {\it long} pulse region where the nonperturbative result 
approaches Schwinger's result (horizontal lines).  
In particular, the deviation is larger for supercritical field 
$|eE_0|/m^2\gtrsim 1$. This can be understood as follows. 
Notice first that the perturbative result always 
approaches 0 in the long pulse limit $m\tau \gg 1$ 
because the typical energies of a (virtual) photon which forms the 
Sauter-type field $\omega \sim 1/\tau \rightarrow 0$ for large $\tau$
and thus not enough to create a pair. On the other hand, Schwinger's
formula valid in the long pulse region says that pair creation 
for subcritical field strength is exponentially suppressed  
$d^3 N/d\mbox{\boldmath $p$}^3 \propto \exp\{ - \pi m^2/|eE_0|  \}$. 
Therefore, the deviation between the two is almost negligible 
for weak field strength $|eE_0|/m^2 \lesssim 1$, while it increases  
with increasing peak strength $|eE_0|/m^2$ in the supercritical 
regime $|eE_0|/m^2 \gtrsim 1$.

The same tendency is found in the comparison of the total number 
of electrons $N$ as shown in Fig.~\ref{fig4}.  We note that the peak 
structure in the {\it short} pulse region is reproduced by the 
perturbative result quite well.  Although 
for small $p_z$ the perturbative value of the density $d^3 N/d\mbox{\boldmath $p$}^3$ is somewhat larger than the nonperturbative one, while it becomes smaller for large $p_z$ (see the left panel of Fig.~\ref{Fig3}),
these differences cancel out with each other in integration over $p_z$.
Thus we have a nice agreement in the total number $N$ as displayed in Fig.~\ref{fig4}.

\begin{figure}[t]
  \begin{center}
   \includegraphics[angle=-90,width=0.7\textwidth]{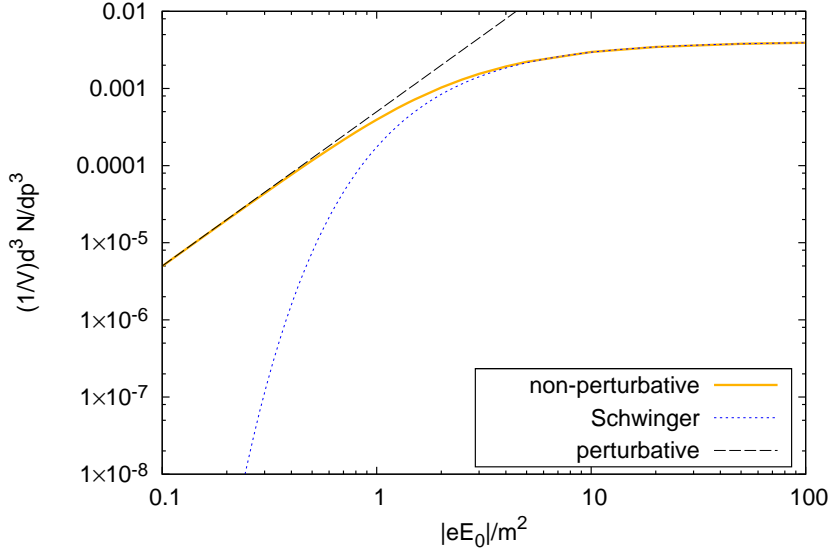}
\caption{(color online).  The peak value 
${\rm max}_{\tau}\left[ d^3 N/d\mbox{\boldmath $p$}^3 \right]$ 
of the nonperturbative result (\ref{eqa16}) (solid line) 
as a function of the field strength $|eE_0|/m^2$. 
For comparison, Schwinger's value and
an estimate 
$(1/V) d^3 N/d\mbox{\boldmath $p$}^3|_{\rm peak} \sim (5.0 \times 10^{-4}) 
\times  (1-p_z^2/p_0^2)|eE_0/p^2_0|^2$ 
obtained with the perturbative result (\ref{eq233})
are shown in dotted and dashed lines, respectively.
Parameters are set to $p_z/m=p_\perp/m=0$.}
  \label{fig7}
 \end{center}
\end{figure}

Figure~\ref{Fig3_taudep} also shows an interesting behavior.  
For relatively short pulses 
$m\tau\lesssim 1$ with subcritical field strength
$|eE_0|/m^2\lesssim 1$ (left panel),
the results of the Sauter-type field are enhanced 
as compared to Schwinger's value (horizontal lines).
Since the pair creation in this region is dominated 
by the perturbative contribution,
this enhancement should be understood as a purely perturbative effect.
It shows up because Schwinger's nonperturbative result 
$d^3 N/d\mbox{\boldmath $p$}^3 \propto \exp\{ - \pi m^2/|eE_0|  \}$
is exponentially small for subcritical field strength 
$|eE_0|/m^2 \lesssim 1$,
while the perturbative result is only power-suppressed 
as $d^3 N/d\mbox{\boldmath $p$}^3 \propto |eE_0/m^2|^2$ (see Eq.~(\ref{eq15})).
By using the perturbative formula for 
$d^3 N/d\mbox{\boldmath $p$}^3$ (\ref{eq233}), we immediately 
find that the peak position $\tau_{\rm peak}$ is given by 
$2 = ( \pi p_0 \tau_{\rm peak} ) \coth[ \pi p_0 \tau_{\rm peak}]$ 
or $ p_0 \tau_{\rm peak} \sim 0.61 $, which does not depend 
on the field strength $|eE_0|$ as is seen in Fig.~\ref{Fig3_taudep}.
Accordingly, the peak value is given by 
$(1/V) d^3 N/d\mbox{\boldmath $p$}^3 \sim (5.0 \times 10^{-4}) 
\times  (1-p_z^2/p_0^2)|eE_0/p^2_0|^2$.

The peak value, 
${\rm max}_{\tau}\left[ d^3 N/d\mbox{\boldmath $p$}^3 \right]$, 
at ${\boldsymbol p}={\bf 0}$
is displayed in Fig.~\ref{fig7} 
as a function of the field strength $|eE_0|/m^2$,
together with Schwinger's value
and the peak value of the perturbative contribution.
The extrapolation of
Schwinger's value to the weak field case $|eE_0|/m^2 \lesssim 1$
underestimates the pair creation in the Sauter-type pulsed field;
the pair creation from the vacuum 
in the region $|eE_0|/m^2 \lesssim 1, m\tau \lesssim 1 $
is actually more abundant than Schwinger's value,
owing to the perturbative contribution with a single photon.
Indeed, the compact formula for the perturbative peak
nicely describes the enhancement for the subcritical fields,
which is
explicitly depicted with a dashed line in Fig.~\ref{fig7}.
Similar behavior was found in Refs.~\cite{SkokovLevai09, sko07},
who however regarded this peak as a result of nonperturbative physics.

\begin{figure}[htbp]
 \begin{center}
  \includegraphics[width=160mm]{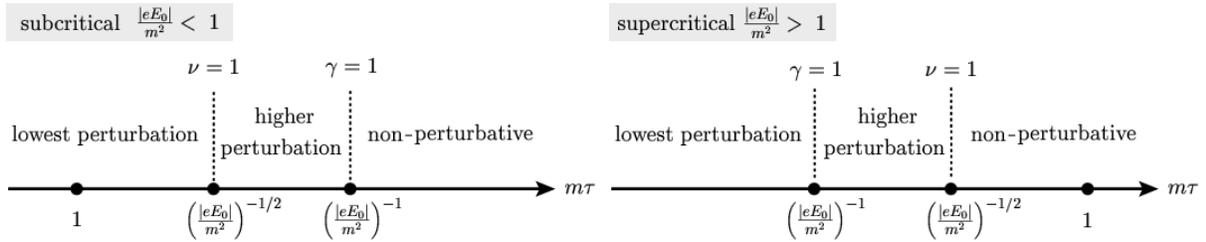}
  \caption{Sketch of the appropriate picture for pair creation from the vacuum for various pulses with height $E_0$ and width $\tau$ .  $m$ is the electron mass.  }
  \label{fig5}
 \end{center}
\end{figure}

Now we return to the question: 
To what extent are we able to say a pulse is {\it short}?  
To answer this, we expand the nonperturbative result (\ref{eqa16}) by the pulse duration $\tau$.  More precisely, we expand ($\ref{eqa16}$) by the following two dimensionless parameters, 
\begin{align}
	\nu \equiv |eE_0| \tau^2 \, ,\\
	\gamma \equiv \frac{|eE_0| \tau}{m}\, ,
\end{align}
because there are two dimensionful quantities $|eE_0|,\, m$ in addition to $\tau$.  The result is 
\begin{align}
	{\rm Eq.(\ref{eqa16})} 
		\! &= \! \frac{  \!\sinh\!\!\left[ \pi \nu \!\left( 1\!+\!\frac{1}{2\gamma} \!\sqrt{\gamma^2 \!-\! 2 \frac{p_z}{p_0} \gamma \!+\! 1  }   \!-\! \frac{1}{2\gamma} \!\sqrt{\gamma^2 \!+\! 2 \frac{p_z}{p_0} \gamma \!+\! 1  }  \right)\! \right]\!  \sinh\!\!\left[\pi \nu \!\left( 1\!-\!\frac{1}{2\gamma} \!\sqrt{\gamma^2 \!-\! 2 \frac{p_z}{p_0} \gamma \!+\! 1  }  \!+\! \frac{1}{2\gamma} \!\sqrt{\gamma^2 \!+\! 2 \frac{p_z}{p_0} \gamma \!+\! 1  }  \right)\! \right]\! }{\sinh\left[\pi\nu \sqrt{1 - 2\frac{p_z}{p_0} \gamma^{-1} + \gamma^{-2}} \right]  \sinh\left[\pi\nu \sqrt{1 + 2\frac{p_z}{p_0} \gamma^{-1} + \gamma^{-2}} \right]} \nonumber\\
		&\longrightarrow \left\{
						\begin{array}{ll}
								\displaystyle
								\frac{1}{(2\pi)^3}\left( 1-\frac{p_z^2}{p_0^2} \right) \left|  
\frac{e E_0}{p^2_0}\right|^2   \frac{(\pi p_0 \tau)^4}{\pi^2 \left|{\rm sinh}[\pi p_0 \tau]\right|^2} 
							&
							\quad	(\nu, \gamma \ll 1) 
							\\
								\displaystyle
								{\rm exp}
									\left[
										-\frac{\pi\left( m^2+\mbox{\boldmath $p$}^2_{\perp} \right)}{|eE_0|}
									\right]
							& 
							\quad	(\nu, \gamma \gg 1)\,.
						\end{array}
					 \right. \label{eq28}
\end{align}
Notice that the asymptotic forms (\ref{eq28}) exactly reproduce 
the perturbative result (\ref{eq233}) for $\nu, \gamma \ll 1 $ and 
the nonperturbative expression for the Schwinger mechanism in a 
constant electric background field \cite{sch51} for $\nu, \gamma \gg 1$.  
Thus, we conclude that pulses, such that the condition 
$\nu, \gamma \ll 1 $ i.e., $ m \tau \ll \sqrt{m^2/|eE_0|}, m^2/|eE_0|$ 
is satisfied, are so {\it short} that pair creation becomes purely 
perturbative, where the {\it lowest} order perturbation theory works 
very nicely.  
On the other hand, pulses, such that the condition $\nu, \gamma \gg 1 $ 
i.e., $ m \tau \gg \sqrt{m^2/|eE_0|}, m^2/|eE_0|$ is satisfied, are 
so {\it long} that pair creation becomes nonperturbative, where 
perturbation theory completely breaks down.  
We can also say that for {\it middle} pulses, such that neither condition $\nu, \gamma \gg 1 $ nor $\nu, \gamma \ll 1 $ is satisfied, perturbation theory is still applicable; however, the lowest-order perturbation theory does not work because higher-order corrections ${\mathcal O}((eE)^n)$ ($n > 1 $) become important.  We summarize our picture in Fig.~\ref{fig5}. 
 These considerations clearly show that in order to investigate the nonperturbative nature of the Schwinger mechanism 
we must require not only the strength $|eE_0|/m^2 \gtrsim 1$ but 
also a sufficient duration $m \tau \gg \sqrt{m^2/|eE_0|}, m^2/|eE_0|$; 
otherwise pair creation from the vacuum can be understood simply as a
perturbative phenomenon.  

The discussion given above is a natural result if we consider
 the meaning of the dimensionless parameters $\nu,\,\gamma$.  
Recall the fact that the work $W$ done by a pulsed electric 
background field with height $E_0$ and width $\tau$ is given 
by $W \sim |eE_0| \tau$ and that the typical energy $\omega$ 
of a photon that forms the pulsed background field is given 
by $\omega \sim 1/\tau$.  Then, we can understand the physical 
meaning of $\nu,\,\gamma$ as follows: 
$\nu \sim W/\omega $ is the number of (virtual) photons of the 
background field involved in a scattering process.  
$\gamma \sim W/m \sim \nu \omega / m$ is the work done by the 
background field scaled by the typical energy scale of the system $m$.  
Keeping these in mind, we can interpret that the perturbative 
condition $\nu, \gamma \ll 1$ corresponds to the case where 
both the number of photons involved in a scattering process $\nu$ 
and its correction to the system $\gamma$ are very small. 
This is obviously a natural criterion for the lowest order 
perturbation theory to work.  We can also interpret the 
nonperturbative condition $\nu, \gamma \gg 1$ in the same way.

It is interesting to compare our discussion with Ref.~\cite{bre70}, which claims that the Keldysh parameter $\gamma^{}_{\rm K} = |eE|/(m\omega)$, where $\omega$ is the typical frequency of the background field, discriminates whether the system is perturbative or nonperturbative.  Note that their discussion is limited to the case where (i) an oscillating electric field background $E(t) = E_0 \cos \omega t$, (ii) $\omega$ is sufficiently small compared to the electron mass $\omega/m \ll 1$, and (iii) the background field is sufficiently weak $|eE_0|/m^2 \ll 1$.  If we assume that the typical frequency $\omega$ of a pulsed background field is given by the inverse of the pulse duration $\omega \sim 1/\tau$, we find that our discussion obtained in a pulsed background field (see Fig.~\ref{fig5}) agrees with Ref.~\cite{bre70} as long as the limitation (iii) is satisfied.  In such a condition, $\gamma$ determines the ``perturbativeness" of the system in our discussion and is equivalent to the Keldysh parameter because $\gamma = \frac{|eE_0|}{m(1/\tau)} \sim |eE_0|/(m\omega) = \gamma_{\rm K}$.

\section{SUMMARY AND DISCUSSION} \label{sec4}

We have explicitly demonstrated that, by using the Sauter-type 
electric field, an interplay between perturbative and nonperturbative 
effects for the $e^+e^-$ pair creation from the time-dependent 
electric field is controlled by two dimensionless parameters 
$\gamma=|eE_0|\tau/m$ and $\nu=|eE_0|\tau^2$. Perturbative pair creation
occurs when $\gamma,\nu\ll 1$ is satisfied, while nonperturbative pair 
creation (the Schwinger mechanism) occurs when $\gamma, \nu\gg 1$ 
is satisfied. In particular, an enhancement of the electron number 
density seen when the pulse duration and the field strength is 
relatively short $m\tau\lesssim 1$ and weak $|eE_0|/m^2 \lesssim 1$, respectively, 
can be understood as the lowest order perturbative process with a single photon.

Throughout this paper, we considered the case where the background field is described by the Sauter-type pulse in order to explicitly perform 
an analytic calculation.  
However, we stress that our qualitative discussion should be valid for more general pulse fields
smoothly characterized by its height $E_0$ and width $\tau$.
It is also interesting to note that, though we focused on the pulse in {\it time} in this paper, our analysis implies that the finite {\it space} extension would also affect the interplay between perturbative and nonperturbative aspects
 of the phenomena under strong fields.

Our analysis is instructive when we consider the effects of 
time-dependent strong fields in actual physical situations. Let us 
briefly discuss the case in high-energy heavy-ion collisions as 
an example. It is estimated that a very strong field $|eE| \gg m^2_e$ 
is generated in heavy-ion collisions operated 
in the Relativistic Heavy Ion Collider (RHIC) at BNL and the Large Hadron 
Collider (LHC) at CERN.  (a) In noncentral collisions such that two nuclei can touch 
each other, numerical simulations\cite{bzd12, den12} have shown 
that its strength is of the order of 
$|eE| \sim 1 \times m_{\pi}^2 \sim (1\times 10^5) \times m_e^2$ 
for RHIC and 
$|eE| \sim 10 \times m_{\pi}^2 \sim (1\times 10^6) \times m_e^2$
for LHC, where $m_\pi$ is the pion mass. 
  (b) In the ultraperipheral collisions where two nuclei do not
touch each other, the electric field is still very strong
and is estimated as $|eE| \sim Z \alpha_{\rm EM} 
\gamma_{\rm L}/b^2$, where $b$ is the impact parameter and $\gamma_{\rm L}$
is the Lorentz factor. With modest parameters, the strength of the field 
reaches $|eE| \sim 60\times m_e^2$ for RHIC 
($Z=79, \gamma_{\rm L} \sim 100, b\sim 1/m_e$) and 
$|eE|\sim (2\times 10^3)\times m_e^2$ for LHC 
($Z=82,\gamma_{\rm L} \sim 3000, b\sim 1/m_e$).  
At first sight, it might be natural to expect 
there exists nonperturbative strong field effects such as the Schwinger
mechanism because the field is extremely strong $|eE|/m_e^2 \gg 1$.
However, from the analysis of the present paper, we have learned 
that we need to be careful about the finite lifetime of the 
strong fields. Indeed, the duration $\tau$ of the strong field is 
extremely short when compared to the typical energy scale of the 
system $m_e$: $m_e \tau \ll 1$.  For
instance, a rough estimate yields  
$\tau \sim 0.1{\rm fm} \sim 3 \times 10^{-4} m_e^{-1}$ (RHIC, LHC) 
for case (a) and 
$\tau \sim b/\gamma_{\rm L} \sim (1\times 10^{-2})\times m_e^{-1}$ 
(RHIC; $\gamma_{\rm L} \sim 100, b\sim 1/m_e$) 
and $\sim 3\times 10^{-4} m_e^{-1}$ 
(LHC; $\gamma_{\rm L} \sim 3000, b\sim 1/m_e$) for case (b).  
With such short durations, the important parameter $\gamma$ can be large $\gamma \gtrsim 1$; however, 
$\nu$ is always so small $\nu \ll 1$ that the pair creation in these processes is 
no longer nonperturbative and the perturbative treatment would be sufficient.  
However, as suggested 
in Ref.~\cite{den12} for case (a), the matter created in the collisions 
could let the electric field survive longer than the naive estimation. 
If this is the case, there is a possibility that pair creation could be
nonperturbative.

\section*{ACKNOWLEDGEMENTS}
This work was supported in part by the Center for the Promotion of
Integrated Sciences (CPIS) of Sokendai and Grants-in-Aid for Scientific Research of MEXT [(C)24540255].

\section*{APPENDIX: ASYMPTOTIC EXPRESSION OF $f(x)$} \label{appA}
Let us find an asymptotic expression of $f(x)$ (\ref{eq222}) for $x \lesssim 1$ and $x \gtrsim 1$.  

For small $x$, we change the variable $\omega$ by $\xi = \omega x$ to obtain
\begin{align}
	f(x) &= \frac{x}{6 \pi^4} \int_x^{\infty} d\xi \xi^2 \sqrt{1-\frac{x^2}{\xi^2}}\left(  2 + \frac{x^2}{\xi^2}  \right) \frac{1}{\sinh^2\xi} \nonumber\\
		 &\sim \frac{x}{6 \pi^4} \int_0^{\infty} d\xi \xi^2  \frac{2}{\sinh^2\xi} \nonumber\\
		 &= \frac{x}{18\pi^2}.  
\end{align}

At large $x$, $\sinh(\omega x) \sim \exp(\omega x)/2$ and only $\omega = 1 + \epsilon \sim 1$ contributes to the integral.  Thus we find
\begin{align}
	f(x) &\sim \frac{2x^4}{3\pi^4} \int_1^{\infty} d\omega \omega^2 \sqrt{1-\frac{1}{\omega^2}} \left( 2 + \frac{1}{\omega^2} \right) {\rm e}^{-2 \omega x} \nonumber\\
		 &= \frac{2x^4}{3\pi^4} \int_0^{\infty} d\epsilon \sqrt{1-\frac{1}{(1+\epsilon)^2}}(2(1+\epsilon)^2 + 1)) {\rm e}^{-2(1+\epsilon)x} \nonumber\\
		 &\sim \frac{2x^4}{3\pi^4} \int_0^{\infty} d\epsilon \left( 3\sqrt{2} \epsilon^{1/2} + \frac{7}{2\sqrt{2}} \epsilon^{3/2} \right) {\rm e}^{-2(1+\epsilon)x} \nonumber\\
		 &= \frac{x^{5/2}}{2\pi^{7/2}} \left( 1+\frac{7}{16} \frac{1}{x} \right) {\rm e}^{-2x}.
\end{align}

\end{document}